\documentclass{iopjournal}
\usepackage[version=3]{mhchem} 
\usepackage{siunitx}
\usepackage{wrapfig}
\usepackage{subcaption}
\usepackage{textcomp}
\newcommand{\hmn}[1]{\ensuremath{\left( #1 \right)}}
\begin{document}

\articletype{Paper}

\title{Substrate-controlled nucleation and growth kinetics in ultrathin Bi$_2$Te$_3$ films}

\author{Damian Brzozowski$^1$\orcid{0000-0003-1631-2044}, Sander R. Hønnås$^1$, Egil Y. Tokle$^1$\orcid{0009-0008-3866-9339}, Jørgen A. Arnesen$^1$ and Ingrid G. Hallsteinsen$^{1,*}$\orcid{0000-0003-0789-8741}}

\affil{$^1$Department of Materials Science and Engineering, Norwegian University of Science and Technology, Trondheim, Norway}

\affil{$^*$Author to whom any correspondence should be addressed.}

\email{ingrid.hallsteinsen@ntnu.no}


\begin{abstract}
Metal chalcogenides are promising layered topological materials, yet their electronic performance is often limited by parasitic bulk conduction arising from defects that introduce excess carriers and shift the Fermi level out of the topological regime. Controlling early-stage growth and defect formation is therefore essential for suppressing bulk transport and enhancing surface-state conduction. Here we investigate ultrathin \ce{Bi2Te3} films grown by pulsed laser deposition on substrates spanning van der Waals, lattice-matched, and amorphous regimes to determine how substrate-dependent nucleation pathways influence defect formation and electronic transport.

Phase-pure, $c$-axis-oriented \ce{Bi2Te3} forms on all substrates, but the growth morphology varies strongly. Layered growth with well-defined quintuple-layer terraces is governed primarily by substrate roughness rather than lattice match: atomically smooth mica and step-terraced \ce{SrTiO3} yield continuous terraces, whereas rougher \ce{BaF2} and amorphous \ce{Si3N4} produce island-structured films. Between the two smooth substrates, the higher surface energy of \ce{SrTiO3} enhances adatom adsorption and nucleation density, promoting rapid vertical growth and early Te depletion.

Transport measurements reveal n-type conduction with carrier densities of $10^{19}$–$10^{20}$ cm$^{-3}$. The highest carrier density occurs for films on \ce{SrTiO3}, consistent with defect formation during high-density nucleation, whereas mobility correlates with structural coherence and terrace formation. Weak anti-localization signatures confirm phase-coherent transport in films on mica and \ce{SrTiO3}. These results show that substrate roughness and nucleation density provide key levers for controlling defect formation and strengthening topological surface transport in \ce{Bi2Te3} thin films.
\end{abstract}

\section{Introduction}
Bismuth telluride (\ce{Bi2Te3}) is a layered van der Waals material with a narrow band gap of \qty{0.15}{eV} and strong spin-orbit coupling. The strong spin-orbit coupling leads to a band inversion between the valence band maximum and the conduction band minimum, giving rise to a topological insulator phase \cite{Zhang2009}. In this state the bulk remains insulating while metallic surface states emerge that are protected by time-reversal symmetry. These spin-momentum-locked surface states have attracted considerable interest for electronic and spintronic application that rely on robust surface transport \cite{Fijalkowski2021,Zhao2024}. In practice, however, electron transport in \ce{Bi2Te3} is often dominated by bulk conduction due to unintentional defects that introduce excess carriers and shift the Fermi level away from the Dirac point. Achieving surface-dominated transport therefore requires precise control over structural and chemical defects that arise during film growth.\newline

Thin films of \ce{Bi2Te3} have been realized using a range of deposition techniques, including molecular beam epitaxy (MBE) \cite{Aabdin2012, Krumrain2011}, pulsed laser deposition (PLD) \cite{Bailini2007, Liao2019}, sputtering \cite{Pandey2023, SHANG2017532}, and physical vapor transport (PVT) \cite{Concepcin2018}. Under optimized conditions these methods typically yield films with strong out-of-plane orientation along the \hmn{003n} crystallographic planes \cite{Li2024, Mogi2017, Tang2022}. These planes correspond to stacks of quintuple layers (QLs) consisting of alternating tellurium and bismuth atomic planes separated by van der Waals gaps. Although such preferential orientation is commonly observed for films ranging from tens of nanometers to micrometers in thickness, the earliest stages of growth can vary significantly in crystalline quality and stoichiometry \cite{Kriegner2017}.\newline

The interaction between the film and the substrate plays a critical role during these initial growth stages. In ideal van der Waals (vdW) epitaxy, layered materials grow on dangling-bond-free surfaces such as mica with minimal lattice constraints. However, when deposited on conventional three-dimensional substrates the interface often exhibits quasi–van der Waals (qvdW) epitaxy, where electrostatic or chemical interactions supplement vdW forces. These interactions influence adatom adsorption, surface diffusion, and nucleation density, thereby affecting terrace formation, grain size, and film morphology \cite{Mortelmans2021, Walsh2017}. Surface roughness and substrate termination further modify these processes by introducing additional nucleation sites and altering local bonding environments.\newline
 
Defects formed during growth strongly influence the resulting electronic properties. Native point defects such as vacancies, antisite defects, and interstitials modify the carrier concentration and shift the Fermi level into bulk bands, thereby suppressing the contribution from topological surface states \cite{Netsou2020, Pereira2021, RANA2024415801}. Structural imperfections such as grain boundaries, mosaicity, and dislocations can also affect electron transport by introducing scattering centers and disrupting crystalline coherence. Transport measurements of \ce{Bi2Te3} films therefore often exhibit contributions from multiple conduction channels. Topological surface states are commonly identified through weak anti-localization (WAL) signatures at low magnetic fields and temperatures. However, WAL may also arise from strong spin-orbit coupling in bulk channels, making it challenging to distinguish surface-related transport from bulk conduction.\newline

Despite extensive studies of \ce{Bi2Te3} thin films, the relationship between substrate-dependent nucleation processes, structural defects, and electronic transport remains incompletely understood. In particular, it is unclear how differences in substrate roughness, bonding environment, and lattice compatibility influence the early stages of film growth and the resulting transport properties.\newline

In this work, we investigate ultrathin \ce{Bi2Te3} films deposited on a set of substrates representing distinct epitaxial environments: muscovite mica enabling vdW epitaxy, \hmn{111}-oriented \ce{BaF2} providing near lattice matching, \hmn{111}-oriented \ce{SrTiO3} representing a qvdW interface, and amorphous \ce{Si3N4} on Si as a non-crystalline surface. By combining structural characterization with electronic transport measurements, we examine how substrate-dependent nucleation and growth pathways influence film morphology, defect formation, and electronic transport in ultrathin \ce{Bi2Te3}. By systematically comparing these different epitaxial environments, this work aims to clarify how substrate roughness, adsorption strength, and lattice compatibility collectively govern the early stages of growth and the resulting electronic behavior. Establishing these relationships provides insight into how growth kinetics shape defect populations and transport properties in layered topological materials, offering guidelines for optimizing thin-film heterostructures for thermoelectric and quantum electronic applications.

\section{Method}
Prior to thin film deposition, each substrate was prepared following dedicated protocols. As-received \ce{Si3N4} substrates (WaferPro) were rinsed in acetone and ethanol. \ce{SrTiO3} substrates (Shinkosha) were etched in a buffered hydrofluoric acid solution (1:9 volumetric ratio of \qty{48}{\%} HF and \qty{40}{\%} \ce{NH4F}) for 45~seconds to ensure titanium single termination and annealed at 1050~°C for 1~hour in an oxygen flow to achieve step terrace surface topography. \ce{BaF2} substrates were annealed at 850~°C for 4~hours in a nitrogen environment. Mica substrates (Agar Scientific) were freshly cleaved with scotch tape prior to deposition to obtain an atomically flat surface.\newline

A KrF excimer laser with wavelength $\lambda$~=~\qty{248}{\nano\meter} was used to ablate a stoichiometric \ce{Bi2Te3} target (Kurt J. Lesker, 99.999$\%$). The target-substrate distance was \SI{50}{\milli\meter}. The PLD chamber remained at a background pressure of \qty{2d-8}{\milli\bar} before introducing argon gas with a working pressure of \qty{1}{\milli\bar}. The substrate temperature was measured with an external pyrometer and set at 220~°C. To minimize deposition growth rate and promote high quality out-of-plane ordering, laser repetition rate was set to \qty{0.2}{\hertz}, and laser fluence to \qty{0.5}{\joule\per\square\centi\meter}. For all substrates, the thickness of the films and the degree of substrate coverage were tuned by changing the number of laser pulses between 10, 25, and 50. Additionally, 1-pulse films were prepared on mica and \ce{SrTiO3} substrates. After deposition, the \ce{Bi2Te3} films were cooled naturally to room temperature in the chamber.\newline

Atomic force microscopy was performed in a Bruker Dimension Icon to analyze the roughness of the films and the degree of surface coverage as a function of laser pulses. The probe used was a ScanAsyst-Air with a nominal tip radius of \qty{2}{\nano\meter}. The Gwyddion software \cite{Necas2012} was used to estimate surface coverage by marking grains by level threshold. Root mean square roughness values were taken on selected topmost layers of deposited films. Crystalline quality was evaluated with Bragg-Brentano X-ray diffraction using Bruker D8 Discover equipped in the copper source (K$\alpha$ $\lambda=\qty{1.5406}{\angstrom}$) and (220)-Ge monochromator. The same system was used to perform X-ray reflectivity to evaluate the thickness of the films. X-ray reflectivity data fitting was performed using the Reflex software and based on the Nelder-Mead simplex algorithm \cite{Vignaud:rg5154}. To confirm proper stoichiometry transfer, Raman spectroscopy was employed. The Renishaw InVia Reflex Spectrometer was operating with the grating of \qty{2400}{\gram\per\milli\meter}, the laser wavelength of \qty{532}{\nano\meter}, and the laser power of \qty{1}{\milli\watt} to prevent surface damage. The signal was averaged over 60 one-second scans.\newline

Magnetoresistance measurements were performed using an attoDRY1000 top-loading pulse tube system with superconducting magnet cryostat. A Keithley 2450 source measure unit, connected to an atto3DR through an ACC100 carrier, was utilized to source currents and measure voltages. Samples were electrically connected through wire bonding with a TPT HB05 Wedge and Ball Bonder in ball/wedge mode, and using silver paint as a conductive adhesive.\newline

For Hall measurements, a current of \qty{10}{\micro\ampere} was applied between two opposite contacts, and the voltage was measured across the remaining pair in the van der Pauw configuration. Weak anti-localization measurements were performed in a standard van der Pauw geometry, with the current sourced between neighboring contacts and the voltage measured across the remaining pair. An external magnetic field was swept symmetrically around zero field between \qty{-1}{\tesla} and \qty{1}{\tesla}. Measurements were performed for both sweep directions and current polarities to suppress thermoelectric offsets and residual Hall contributions.
To analyze weak anti-localization, low-field magnetoresistance curves were recorded at temperatures between \qty{5}{\kelvin} and \qty{50}{\kelvin}. The weak anti-localization features were fitted using the Hikami–Larkin–Nagaoka model to extract the $\alpha$ prefactor and the phase-coherence length $l_\phi$.

\section{Results}
\subsection{Substrate-dependent growth and morphology of \ce{Bi2Te3}}
To investigate the early stages of film growth, a thickness series corresponding to 10, 25, and 50 laser pulses was deposited on all four substrates. In addition, 1-pulse films were prepared on mica and \ce{SrTiO3} to probe the earliest nucleation stages. All other deposition parameters were kept constant.\newline

\subsubsection{Phase formation and stoichiometry.}
\begin{figure}[h]
  \centering
  \includegraphics[width=0.5\linewidth]{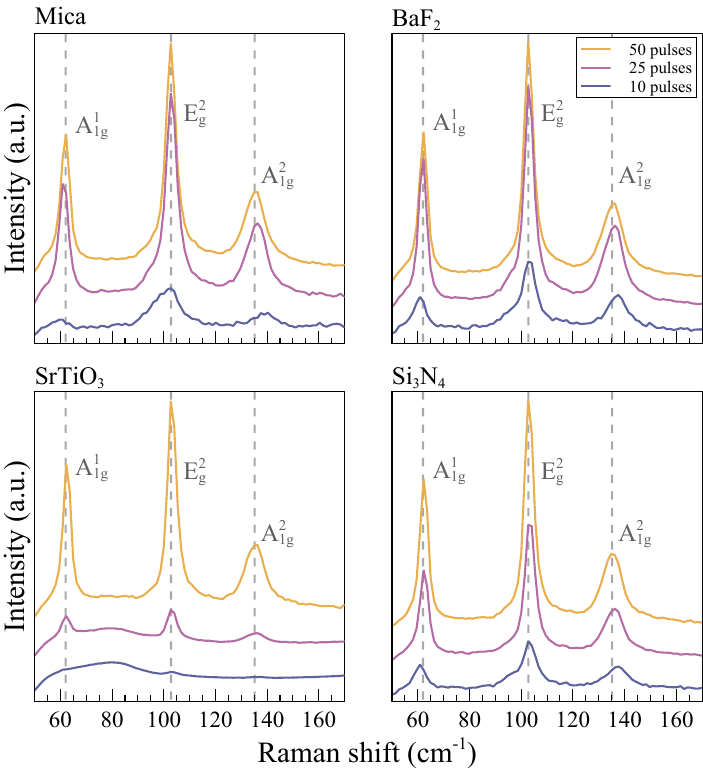}
  \caption{Raman spectra of deposited films. Dashed lines indicate characteristic phonon modes of \ce{Bi2Te3}.}
  \label{fig1}
\end{figure}

Raman spectroscopy was used to verify the stoichiometry of the deposited films through their characteristic vibrational modes \cite{Concepcin2018}. The Raman spectra of the films are shown in Figure~\ref{fig1}. For the 50-pulse films (yellow lines), peaks are observed at Raman shifts of 62, 103, and approximately 135~cm$^{-1}$, corresponding to the A$^{1}_{1g}$, E$^{2}_{g}$ and A$^{2}_{1g}$ vibrational modes of \ce{Bi2Te3}. The presence of these modes confirms that the deposited films retain the characteristic 2:3 stoichiometry during laser ablation and adsorption \cite{Concepcin2018,Russo2008,KUZNETSOV2016122,Xu2015Raman}.\newline

The Raman intensity decreases for the 25- and 10-pulse (purple and blue lines), consistent with the reduced thickness of these samples. In addition, the A$^{2}_{1g}$ mode shifts from approximately 133~cm$^{-1}$ for the 50-pulse films to about 135~cm$^{-1}$ and 138~cm$^{-1}$ for the 25- and 10-pulse films, respectively. This shift is attributed to phonon confinement effects that become increasingly pronounced in ultrathin \ce{Bi2Te3} layers, consistent with previous reports for discontinuous films composed of isolated grains \cite{rodrigues2023}. In addition, the relative ratio of E$^{2}_{g}$ to A$_{1g}$ is reduced with decreasing thickness, which is a result of the alleviated constraining of the out-of-plane A$_{1g}$ phonon modes in thinner films \cite{Shahil2010}. The calculated intensity ratios are provided in the Supporting Information.\newline

For the 10-pulse film deposited on \ce{SrTiO3}, the characteristic \ce{Bi2Te3} Raman modes are absent and replaced by a broad feature around 80~cm$^{-1}$. Such a signal has been associated with tellurium-deficient phases such as \ce{BiTe} and \ce{Bi4Te3} \cite{Concepcin2018}. This behavior may arise from selective absorption or desorption of Bi and Te adatoms on the polar \ce{SrTiO3} \hmn{111} surface during the earliest growth stage. As the film thickness increases, the characteristic \ce{Bi2Te3} modes reappear, indicating that stoichiometric growth is restored.\newline

\subsubsection{Crystallographic orientation and growth rate.}
\begin{figure}
  \centering
  \includegraphics[width=0.9\textwidth]{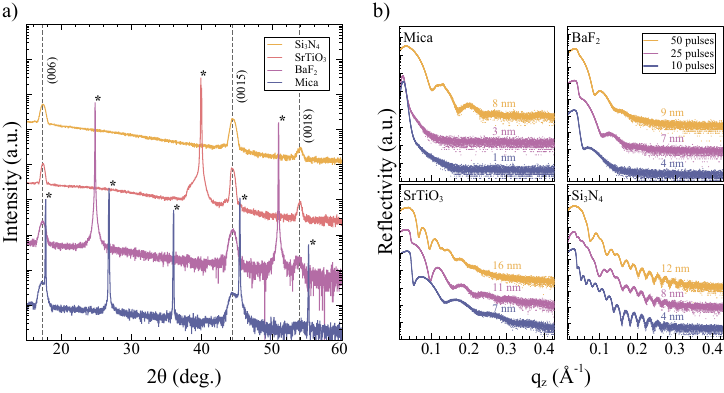}
  \caption{a) XRD wide-scans of the 50 pulse samples, plotted in logarithmic scale. Substrate peaks marked with asterisks. The dashed lines mark \hmn{003n} peaks of \ce{Bi2Te3}. b) XRR profiles of the samples. The profiles were used to estimate sample's thickness. The film grown on \ce{Si3N4} displays additional, high-frequent reflectivity curves from the \ce{Si3N4}/Si interface}
  \label{fig2}
\end{figure}

X-ray diffraction (XRD) and X-ray reflectivity (XRR) were used to examine the crystallographic orientation and thickness of the films. Widescan Bragg-Brentano diffractograms of the 50-pulse samples are shown in Figure~\ref{fig2}(a). Additional diffractograms of the 25- and 10-pulse samples are included in the Supporting Information. All films exhibit reflections at $2\theta$ values corresponding to the \hmn{006}, \hmn{0015} and \hmn{0018} lattice planes as marked with the dashed lines. No additional reflections are observed, indicating that the films grow with a strong out-of-plane preferential orientation along the \hmn{003n} family of planes regardless of substrate type.\newline

XRR measurements reveal substantial differences in film thickness for samples deposited with the same number of laser pulses, as shown in Figure~\ref{fig2}(b). For example, the films grown with 50 pulses on mica reaches a thickness of approximately 8~nm, whereas the corresponding film on \ce{SrTiO3} reaches approximately 16~nm. Similar differences are observed for the 25- and 10-pulse films. The 10-pulse film on \ce{SrTiO3} already reaches a thickness of about 7~nm, indicating rapid out-of-plane growth. In contrast, films on \ce{Si3N4} and \ce{BaF2} reach only 4~nm. On mica, the growth is significantly slower in the vertical direction, reaching only about 1~nm after 10 pulses - atomic force microscopy (AFM) scans included in the Supporting Information confirm these thickness estimations.\newline

\begin{figure}[h]
  \centering
  \includegraphics[width=1\linewidth]{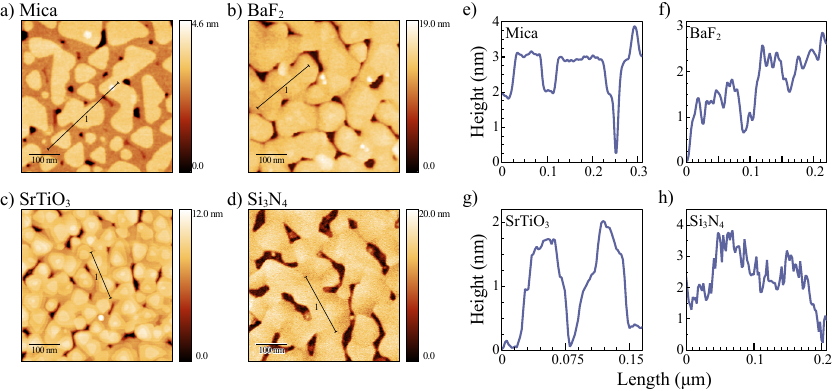}
  \caption{Left panel: 500$\times$500~nm AFM scans of 50 pulse samples. Right panel: height profiles extracted from AFM scans.}
  \label{fig3}
\end{figure}

\subsubsection{Surface morphology.}
The surface morphology of the 50-pulse films was examined using AFM, as shown in Figure~\ref{fig3}. The film grown on mica, Figure~\ref{fig3}(a), exhibits a layered morphology with well-defined terraces corresponding to individual QLs. The surface coverage reaches approximately \qty{98}{\%}, resulting in a nearly continuous film with low porosity. Triangular grains with two in-plane orientations rotated by 60° are observed in the upper layer, with an average grain size of approximately \qty{100}{\nano\meter}. The lower film layer does not show clear grain boundaries, indicating that the grains are in-plane compatible and eventually coalesce. Step-like features and triangular grains are indicative of the high-quality layer-by-layer deposition \cite{Liu2012, Ferhat2000}. The root-mean-square (RMS) roughness is \qty{0.7}{\nano\meter}, and the extracted step height between adjacent terraces, Figure~\ref{fig3}(e), is approximately \qty{1}{\nano\meter}, consistent with the thickness of a single QL.\newline

The film grown on \ce{SrTiO3}, Figure \ref{fig3}(c), also exhibits a layered morphology with low roughness (RMS \qty{1.3}{\nano\meter}) and high surface coverage of approximately 95\%. However, the grains are smaller, typically \qtyrange{50}{80}{\nano\meter} in size, and multiple stacked layers are visible, indicating a stronger contribution from out-of-plane growth compared to the mica substrate. The height profile shown in Figure~\ref{fig3}(g) is not as well defined as that of the mica-based film. However, the step height corresponds closely to the height of 1 QL.\newline

In contrast, films deposited on \ce{BaF2} and \ce{Si3N4}, Figure~\ref{fig3}(b,d), display a markedly different morphology, which is reflected in irregular height profiles plotted in Figure~\ref{fig3}(f,h). Instead of well-defined terraces, these films form coalesced networks with irregular grain boundaries and higher porosity. The RMS roughness is significantly larger, particularly for \ce{Si3N4}, where values of up to \qty{3.4}{\nano\meter} are observed. The scans show no clear separation between the out-of-plane layers or in-plane alignment. The grain sizes range from \qtyrange{100}{200}{\nano\meter}.\newline

Additional comparison of the surface coverage is continued based on the scanning electron microscopy (SEM) images of the 10-pulse films. The images are used to extract the coverage percentage and are included in the Supporting Information. As the thickness decreases and films become progressively less continuous, isolated grain features become dominant on all substrates. The surface coverage of the 10-pulse films varies significantly depending on the substrate type. The film grown on \ce{SrTiO3} reaches approximately 87$\%$ coverage, whereas the coverage is about 80\% on mica, 76\% on \ce{BaF2}, and approximately 60\% on \ce{Si3N4}. This is indicative of a changed deposition rate relative to the 50-pulse films, where the mica-based film yields the highest coverage.\newline

\subsubsection{Role of substrate surface properties.}
The observed differences in film morphology can be related to the surface characteristics of the substrates. Supplementary AFM measurements of the bare substrates show that amorphous \ce{Si3N4} possesses the highest surface roughness. Both \ce{BaF2} and \ce{SrTiO3} exhibit step-terraced surfaces, although the step-edges of \ce{BaF2} contain protruding particulates that are absent on \ce{SrTiO3}. In contrast, freshly cleaved mica provides an atomically flat surface. These variations in surface morphology strongly influence the early stages of film growth. Films grown the smoother substrates (mica and \ce{SrTiO3}) develop layered morphologies with relatively uniform grain distributions, whereas films grown on the rougher substrates (\ce{BaF2} and \ce{Si3N4}) form disordered networks with higher roughness and reduced terrace formation. This comparison indicates that substrate roughness plays a critical role in determining the growth kinetics and morphology of ultrathin \ce{Bi2Te3} films during the early stages of deposition.\newline

\subsection{Nucleation}
To further probe the earlier stages of film formation, 1-pulse depositions of \ce{Bi2Te3} were performed on mica and \ce{SrTiO3}. Nucleation analysis was not carried out on \ce{BaF2} and \ce{Si3N4}. As discussed in the previous section, both substrates exhibit substantial surface roughness at the nanometer scale, which makes it difficult to reliably distinguish newly formed nuclei from the underlying topography in AFM scans. In contrast, mica provides an atomically flat surface upon cleavage, while \ce{SrTiO3} exhibits well-defined step-terrace structures. These smoother surfaces enable reliable identification of individual nuclei and therefore allow a quantitative analysis of nucleation density.\newline

\begin{figure}[h]
  \centering
  \includegraphics[width=0.7\linewidth]{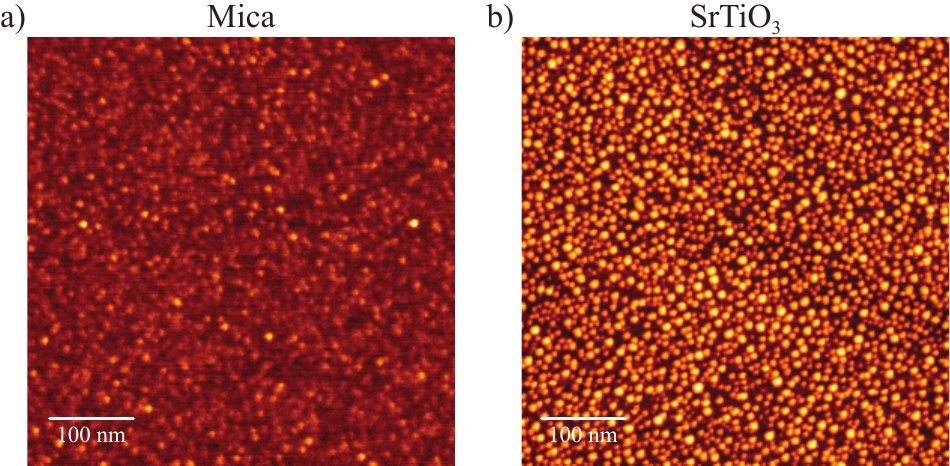}
  \caption{500$\times$500~nm AFM scans of 1-pulse films grown on mica (a) and \ce{SrTiO3} (b) substrates. The background level is scaled to highlight the bright-colored nucleation sites at the surface.}
  \label{fig4}
\end{figure}

Figure~\ref{fig4} shows AFM images of the 1-pulse samples on mica and \ce{SrTiO3}. On \ce{SrTiO3}, the surface is populated by small grains with lateral dimensions of approximately \qtyrange{5}{10}{\nano\meter}. AFM line profiles indicate heights between \qtyrange{2}{3}{\nano\meter}, corresponding to roughly three QLs. The resulting surface coverage is approximately 55$\%$. In contrast, the film grown on mica exhibits significantly lower coverage of about 25$\%$, with islands of approximately \qty{1}{\nano\meter} height corresponding to a single QL. These values translate into a substantial difference in the amount of adsorbed material per area, corresponding to approximately 10 formula units per nm$^2$ on \ce{SrTiO3} and about 1.5 formula units per nm$^2$ on mica. Because both samples were deposited under identical conditions and therefore experience identical plume dynamics, the observed differences arise from the interaction between incoming species and the substrate surface. The reduced adsorption on mica is consistent with the nature of vdW substrates, which lack dangling bonds and therefore interact only weakly with incoming adatoms. In contrast, the \ce{SrTiO3} surface exposes partially charged and chemically active terminations that provide stronger adsorption sites.\newline

To quantify the nucleation process, we employ the framework described by Mortelmans et al. for layered dichalcogenides \cite{Mortelmans2020}. Their work established a relationship between nucleation density and defect formation during layered material growth. In particular, higher nucleation densities are often associated with increased densities of structural defects such as grain boundaries and stacking faults. The model also allows comparison of nucleation densities across different growth techniques and substrates by normalizing values to a coverage fraction of 30\% \cite{Mortelmans2021}. The approach assumes a critical island size of one atom, negligible desorption, and isotropic two-dimensional diffusion. 
The normalized nucleation density, $N_{30\%}$, is given by
\begin{equation}
    N_{30\%} = \sqrt[3]{\frac{3\theta^2_{30\%}}{t_{30\%}D_{30\%}}}
    = N \cdot \sqrt[3]{\frac{0.3}{\theta \cdot a^2_{\textrm{Bi}_2\textrm{Te}_3} \cdot \sqrt{3}/2}} \ ,
\end{equation}
where $N$ is the nucleation density, $\theta$ is the surface coverage, $t$ is the growth time, $D$ is the diffusion constant, $a_{\textrm{Bi}_2\textrm{Te}_3}$ is the in-plane lattice parameter of \ce{Bi2Te3}, and the subscript 30\% denotes normalization to 30\% coverage \cite{Mortelmans2021}. This expression can be reformulated in terms of quantities directly obtained from AFM measurements as 
\begin{equation}
    N_{30\%} = N \cdot \sqrt[3]{\frac{0.3}{\varphi} \cdot \frac{c_{\textrm{Bi}_2\textrm{Te}_3}}{3 h}} \ ,
\end{equation}
where $c_{\textrm{Bi}_2\textrm{Te}_3}$ is the out-of-plane lattice parameter of the \ce{Bi2Te3} unit cell, $\varphi$ is the coverage fraction, and $h$ is the film thickness \cite{Mortelmans2020}. The parameters extracted from the AFM analysis are summarized in Table~\ref{tab1}.\newline

\begin{table}
  \caption{Nucleation density $N$, thickness $h$, coverage fraction $\varphi$, and normalized nucleation density $N_{30\%}$ of the one-pulse films}
  \centering
  \label{tab1}
    \begin{tabular}{l|l|l|l|l}
    \hline
    1 pulse & $N$ (cm$^{-2}$) & $h$ (\qty{}{\nano\meter})  & $\varphi$ ($\%$ area)  & $N_{30\%}$ (cm$^{-2}$) \\
    \hline 
    Mica  & \qty{5e11}{} & 1 & 25 & \qty{7e11}{}  \\
    SrTiO$_{3}$  & \qty{9e11}{} & 3 & 56 & \qty{5e11}{}\\
    \hline
  \end{tabular}
\end{table}

\begin{figure}
  \centering
  \includegraphics[width=0.7\linewidth]{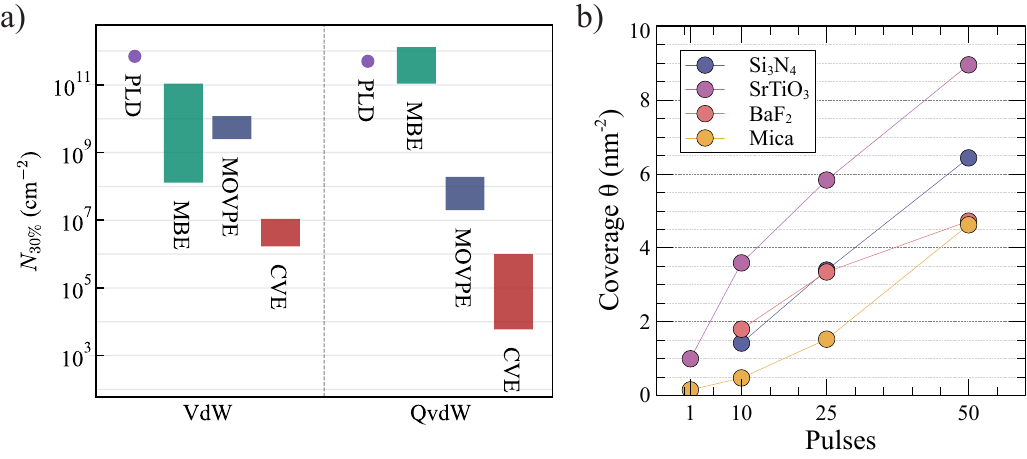}
  \caption{a) Comparison of normalized nucleation density between different growth techniques for van der Waals epitaxy and quasi van der Waals epitaxy. The ranges for MBE, MOVPE, and CVE are taken from \cite{Mortelmans2020}, while the PLD values are calculated based on the 1-pulse samples. b) Calculated coverage $\theta$ values for all samples discussed.}
  \label{fig5}
\end{figure}

Figure~\ref{fig5}(a) presents the resulting $N_{30\%}$ values for the one-pulse films on mica and \ce{SrTiO3}, together with the literature ranges for other growth techniques compiled in \cite{Mortelmans2020}. The calculated normalized nucleation densities are \qty{5e11}{} nuclei per cm$^2$ for \ce{SrTiO3} and \qty{7e11}{} nuclei per cm$^2$ for mica. These values fall within the range typically reported for qvdW epitaxial growth by molecular beam epitaxy (MBE), while exceeding most reported values for purely vdW epitaxy across different deposition techniques. The relatively high nucleation densities observed for PLD are consistent with its supersaturated growth environment and comparatively low growth temperature. In the present study, \ce{Bi2Te3} growth is carried out at \qty{220}{\celsius}, which is lower than temperatures typically employed by MBE or vapor-phase techniques and therefore limits adatom diffusion.\newline

Using the measured surface coverage $\theta$, the growth rate of the films deposited with 10, 25, and 50 pulses can also be estimated. The resulting values are summarized in Figure~\ref{fig5}(b). A clear substrate-dependent trend emerges. On mica, the growth rate is initially lower, reflecting the reduced adsorption probability of the vdW surface. As the \ce{Bi2Te3} film begins to cover the substrate, the growth rate increases as incoming species increasingly interact with previously deposited \ce{Bi2Te3}. In contrast, films grown on \ce{SrTiO3} exhibit the opposite behavior: the initial growth rate is relatively high but decreases with increasing thickness as the influence of the substrate surface diminishes and adatom interactions become dominated by the growing film itself. At larger thicknesses the growth rates across different substrates converge, as also observed for films grown on \ce{Si3N4}. The films deposited on \ce{BaF2} constitute an outlier in this analysis, showing an apparent stagnation of growth relative to the other substrates, and are therefore excluded from further discussion.\newline

Overall, the results indicate that \ce{Bi2Te3} nucleation densities during PLD are comparable to the observed for MBE growth, suggesting that nucleation is primarily determined by deposition conditions rather than the epitaxial regime alone. However, the substrate strongly affects the subsequent growth kinetics. The higher nucleation density observed on \ce{SrTiO3} does not necessarily promote two-dimensional growth. Instead, the reduced diffusion length leads to rapid nucleation of many small islands, which subsequently grow preferentially out-of-plane once quintuple-layer nuclei are established. In contrast, the weaker adsorption on mica allows adatoms to diffuse over longer distances, resulting in fewer nuclei and more pronounced lateral island expansion. As growth proceeds and grains coalesce, the morphology of films on both substrates converges toward faceted domains with three-fold symmetry and well-defined QL steps.

\subsection{Electronic transport}
To investigate how structural and point defects influence the functional properties of the films, their electronic transport was characterized using van der Pauw measurements. Table~\ref{tab2} summarizes the transport parameters measured at \qty{5}{\kelvin} for \ce{Bi2Te3} films deposited with 50 laser pulses on the different substrates. All samples exhibit n-type conduction, indicating that the Fermi level lies within the conduction band. This behavior is consistent with slight Te deficiency, where tellurium vacancies act as donor-like defects and dominate the bulk charge transport.\newline

\begin{table}
  \centering
  \caption{Electronic transport properties at \qty{5}{\kelvin} of \ce{Bi2Te3} deposited with 50 pulses on selected substrates. Table includes resistivity, conductivity, carrier density and mobility}
  \label{tab2}
  \begin{tabular}{l||l|l||l||l}
    \hline
    & $\rho$ (\si{\ohm\metre})
    & $\sigma$ (\si{\siemens\per\metre})
    & $n_{3D}$ (\si{cm^{-3}})
    & $\mu$ (\si{cm^{2}\per\volt\second}) \\
    \hline\hline
    \textit{(001)Mica} & $1.693\cdot10^{-5}$ & $5.908\cdot10^4$ & $4.402\cdot10^{19}$ & 83.76 \\
    \textit{(111)BaF$_2$} & $2.428\cdot10^{-5}$ & $4.118\cdot10^4$ & $1.184\cdot10^{20}$ & 21.70 \\
    \textit{(111)SrTiO$_3$} & $1.520\cdot10^{-5}$ & $6.579\cdot10^4$ & $1.540\cdot10^{20}$ & 26.66 \\
    \textit{Si$_3$N$_4$} & $1.720\cdot10^{-4}$ & $5.813\cdot10^3$ & $1.021\cdot10^{20}$ & 3.55 \\
    \hline
  \end{tabular}
\end{table}

Hall measurements reveal relatively high carrier concentrations in the range of $10^{19}$ to $10^{20}$~cm$^{-3}$, placing all films firmly within the bulk-conducting regime. Under these conditions, topological surface states are unlikely to dominate total conductance, although they may still dominate the phase-coherent component of the transport \cite{analytis2010bulk}. The measured mobility and resistivity values fall within the range typically reported for \ce{Bi2Te3} films grown by PLD and measured in the van der Pauw geometry \cite{Tang2022}. However, compared to mobility values extracted from Shubnikov–de Haas measurements, the values observed here are relatively low, suggesting substantial carrier scattering in all films \cite{ngabonziza2022quantum}.

Clear substrate-dependent trends are observed when comparing the films. The film grown on mica exhibits the lowest resistivity (\qty{1.693e-5}{\ohm\meter}), the highest mobility (83.76 \si{cm^{2}\per\volt\second}), and the lowest carrier density \qty{4.4e19}{cm^{-3}}. These values are generally considered favorable for accessing topological transport in \ce{Bi2Te3}. Films grown on \ce{SrTiO3} and \ce{BaF2} display similar resistivity values of \qty{1.5e-5}{\ohm\meter} and \qty{2.4}{\ohm\meter}, but substantially reduced mobility of \qty{21.7}{cm^2/Vs} and \qty{16.7}{cm^2/Vs}, respectively, indicating increased carrier scattering. The film deposited on \ce{Si3N4} exhibits the most degraded transport performance, with both resistivity and mobility differing by approximately an order of magnitude from those measured for the mica sample. This strongly suggests that structural disorder and porosity dominate the transport behavior and lead to defect-induced scattering.\newline

The observed mobility trends correlate well with the structural characterization presented in the previous sections. Films grown on mica exhibit the most ordered morphology with large, well-coalesced terraces and minimal porosity, which reduces scattering at grain boundaries. In contrast, the films grown on \ce{Si3N4} display the most disordered morphology, characterized by rough surfaces and a porous island network that significantly suppresses carrier mobility. Films grown on \ce{SrTiO3} and \ce{BaF2} exhibit intermediate mobilities, consistent with their intermediate topographical quality roughness associated with the step-terrace structure. These results suggest that grain boundaries and structural disorder play a dominant role in limiting carrier mobility.\newline

Carrier density follows a different trend. The highest carrier concentration is observed for the film grown on \ce{SrTiO3}, reaching \qty{1.54e20}{cm^{-3}}, while the films on \ce{BaF2} and \ce{Si3N4} exhibit similar values near \qty{1.2e20}{cm^{-3}} and \qty{1.0e20}{cm^{-3}}, respectively. The increased carrier density in the \ce{SrTiO3} film may be related to increased point-defect formation during the earliest stages of growth. This interpretation is consistent with the Raman analysis of the 10-pulse films, which indicated deviations from the ideal stoichiometry on this substrate.\newline

While the Hall measurements indicate substantial bulk conduction, magnetotransport measurements provide a sensitive probe of the phase-coherent contribution to the transport. In particular, weak anti-localization arises from quantum interference effects in systems with strong spin-orbit coupling. In such systems, time-reversed electron path interfere destructively, suppressing backscattering and producing a characteristic cusp in the magnetoresistance near zero magnetic field \cite{lu2014weak}. WAL is frequently observed in topological insulators where spin-orbit coupling is strong and phase coherence can persist over relatively long length scales \cite{he2011impurity}.\newline

In two-dimensional systems the WAL correction to the classical Drude conductivity is described by the Hikami–Larkin–Nagaoka (HLN) model:
\begin{equation}
\label{HLN_Model}
\Delta \sigma(B) = -\alpha \frac{e^2}{2\pi^2\hbar} \left[ \psi\left(\frac{1}{2} + \frac{B_\varphi}{B} \right) - \ln\left( \frac{B_\varphi}{B} \right) \right],
\end{equation}
where $\Delta \sigma(B)$ is the change in conductivity as a function of magnetic field $B$, $\alpha$ is a dimensionless prefactor related to the number of coherent conduction channels, $\psi$ is the digamma function \cite{abramowitz1965formulas}, and $B_\varphi$ is the characteristic dephasing field \cite{hikami1980spin}. The dephasing field is related to the phase-coherence length $l_\varphi$ through
\begin{equation}
B_\varphi = \frac{\hbar}{4 e l_\varphi^2}.
\end{equation}
Since the HLN model is formulated in terms of magnetoconductivity, the experimentally measured resistivity $\rho(B)$ is converted to conductivity using the approximate film thickness $t$ according to
\begin{equation}
\sigma(B) = \frac{t}{\rho(B)}.
\end{equation}
The conductivity correction used for fitting is then
\begin{equation}
\Delta \sigma(B) = \sigma(B) - \sigma(0).
\end{equation}

\begin{figure}
  \centering
  \includegraphics[width=\linewidth]{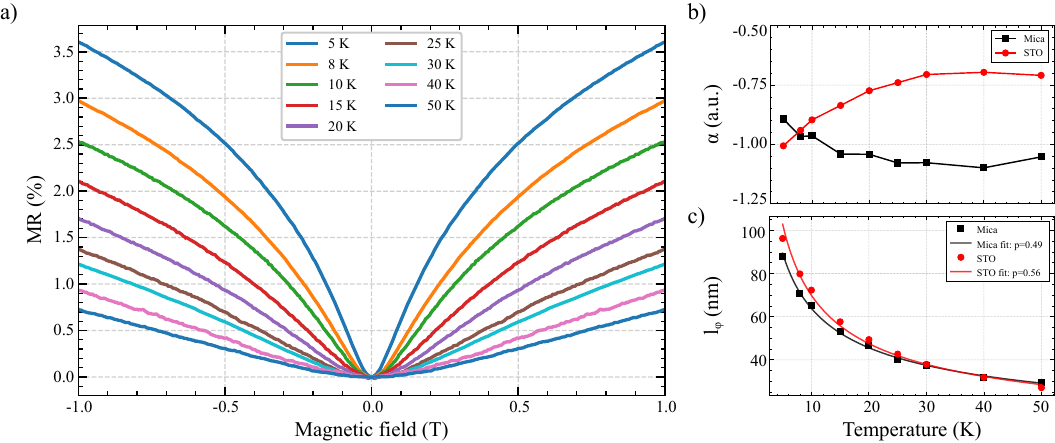}
  \caption{a) Magnetoresistance (MR$\%$) of the \ce{Bi2Te3} film grown on mica showing clear WAL cusps that weaken with increasing temperature. b) HLN prefactor $\alpha$ for the mica and \ce{SrTiO3} films. c) Phase-coherence length $l_\phi$ with power-law fits $l_\phi \propto T^{-p/2}$.}
  \label{fig6}
\end{figure}

Magnetoresistance measurements were performed on the films grown on mica and \ce{SrTiO3}. Both samples exhibit pronounced WAL cusps at low magnetic fields, characteristic of strong spin–orbit coupling. Figure~\ref{fig6}(a) shows the normalized magnetoresistance for the mica sample (data for \ce{SrTiO3} are shown in the Supporting Information). The WAL feature gradually weakens with increasing temperature. The extracted HLN parameters for both films are summarized in Figure~\ref{fig6}(b,c), which shows the temperature dependence of the prefactor $\alpha$ and the phase-coherence length ($l_\phi$).\newline

For a single two-dimensional conduction channel with strong spin-orbit coupling the HLN model predicts $\alpha=-\frac{1}{2}$. Additional coherent channels increase the magnitude of $\alpha$. Contributions from bulk conductivity or disorder-induced electron-electron interactions cause the prefactor to deviate toward positive values \cite{chiu2013weak}. For the mica sample, the prefactor remains close to $\alpha\approx-1$ across the entire temperature range. This value is often associated with two coherent and decoupled topological surface states \cite{steinberg2011electrically, chen2010gate}. However, given the relatively high carrier density of $n_{3D}=4.4\cdot10^{19}$, a coherent bulk contribution is also expected \cite{analytis2010bulk}. In this case the observed $\alpha\approx-1$ likely reflects the combined contribution of a partially coherent bulk channel and a surface-derived channel, rather than two isolated surface states \cite{PhysRevLett.109.066803}.\newline

The film grown on \ce{SrTiO3} exhibits a similar low-temperature value of $\alpha\approx-1$ at \qty{5}{\kelvin}, but the prefactor gradually approaches $\alpha\approx-0.7$ as the temperature increases toward \qtyrange{30}{50}{\kelvin}. This behavior suggests that the surface contribution gradually loses phase coherence with increasing temperature, while the bulk channel remains relatively robust, consistent with previous observations in topological materials \cite{lu2014weak}.\newline

The temperature dependence of the phase-coherence length further clarifies the dominant dephasing mechanism \cite{barthel2008dephasing}. For both samples the phase-coherence length follows a power-law dependence \cite{lin2002recent} of the form
\begin{equation}
    l_\phi\propto T^{-p/2}.
\end{equation}
Fitting the experimental data yields $p=0.49$ for the mica film and $p=0.56$ for the \ce{SrTiO3} film, values characteristic of electron-electron interaction dominated dephasing in two-dimensional diffusive systems \cite{efros2012electron}. This behavior is consistent with a significant contribution of two-dimensional conduction channels to the phase-coherent transport.\newline

The combined structural and electronic measurements reveal a clear hierarchy of mechanisms that govern the transport properties of ultrathin \ce{Bi2Te3} films. Substrate roughness first determines the effective diffusion landscape for incoming adatoms, thereby controlling nucleation density and island coalescence during the earliest stages of growth. These nucleation conditions subsequently determine grain size, terrace formation, and structural coherence of the films. Structural coherence primarily governs electron mobility through scattering at grain boundaries and morphological disorder. In parallel, substrate surface chemistry influences adsorption strength and early growth stoichiometry, which affects the formation of point defects that control the carrier density. Within this framework, atomically flat substrates such as mica promote extended adatom diffusion, lower nucleation density, and terrace-forming growth, resulting in higher mobilities and lower carrier concentrations. In contrast, substrates with higher surface reactivity or roughness increase nucleation density and defect formation, which suppress mobility and increase bulk carrier density. Notably, these results demonstrate that substrate roughness and substrate-dependent growth kinetics can dominate over lattice mismatch in determining the structural and electronic quality of PLD-grown \ce{Bi2Te3} films.

\section{Conclusion}
We systematically investigated the early-stage growth and electronic transport of ultrathin \ce{Bi2Te3} films deposited by pulsed laser deposition on substrates spanning distinct epitaxial regimes. While phase-pure, $c$-axis-oriented \ce{Bi2Te3} films were obtained on all substrates, the surface morphology and growth pathways varied significantly. Atomically flat mica and step-terraced \ce{SrTiO3} promote layered growth with well-defined quintuple-layer terraces, whereas films grown on \ce{BaF2} and amorphous \ce{Si3N4} exhibit more irregular, island-structured morphologies. These differences correlate primarily with substrate roughness rather than nominal lattice mismatch.\newline 

Analysis of the earliest growth stages reveals substrate-dependent kinetic pathways. On mica, weak van der Waals interactions reduce adatom adsorption and favor the formation of single-quintuple-layer nuclei that expand predominantly in-plane. In contrast, the more reactive \ce{SrTiO3} surface enhances adsorption, leading to higher nucleation densities and the formation of multi-QL nuclei that promote faster out-of-plane growth. Comparison with literature values shows that the normalized nucleation densities obtained by PLD are comparable to those reported for molecular beam epitaxy.\newline

Transport measurements reveal n-type conduction across all samples with carrier densities in the range of $10^{19}$–$10^{20}$~cm$^{-3}$, consistent with defect-induced doping dominated by tellurium vacancies. Electron mobility correlates strongly with structural coherence: films exhibiting terrace-forming growth on smooth substrates show higher mobilities, whereas rough or porous morphologies suppress mobility through enhanced scattering at grain boundaries and structural disorder. Magnetotransport measurements reveal clear weak anti-localization signatures in the films grown on mica and \ce{SrTiO3}, indicating the presence of phase-coherent conduction channels associated with strong spin–orbit coupling. Temperature-dependent measurements further suggest that surface-state contributions remain observable up to \qty{50}{K} for the mica films, while they are increasingly masked by bulk conduction in the \ce{SrTiO3} samples due to their higher carrier densities.\newline

Taken together, the structural and electronic measurements reveal a hierarchy of mechanisms linking substrate properties to electronic transport in ultrathin \ce{Bi2Te3}. Substrate roughness and surface chemistry determine adatom diffusion and nucleation density during the earliest stages of growth, which subsequently control grain size, terrace formation, and defect density. These structural characteristics ultimately govern carrier mobility and bulk carrier concentration. The results demonstrate that substrate-controlled growth kinetics—particularly surface roughness—can dominate over lattice matching in determining the structural and electronic quality of PLD-grown \ce{Bi2Te3} films. By identifying nucleation density and substrate morphology as key parameters, this work provides practical design guidelines for engineering \ce{Bi2Te3} heterostructures with reduced defect densities and enhanced topological surface transport.

\funding{This project was supported by the Research Council of Norway under Project Number 325063. The Research Council of Norway is acknowledged for the support to the Norwegian Micro- and Nano-Fabrication Facility, NorFab, project number 295864.}

\roles{D.B.: Conceptualization, data curation, formal analysis, investigation, methodology. S.R.H.: Data curation, formal analysis, investigation. E.Y.T.: Data curation, formal analysis, investigation. J.A.A.: Data curation, formal analysis, investigation. I.G.H.: Conceptualization, funding acquisition, methodology, project administration.}

\data{All data that support the findings of this study are included within the article (and any supplementary files).}

\suppdata{The following files are available free of charge. Supporting information containing: Raman intensity ratios, diffractograms and AFM scans of 10- and 25-pulse samples, SEM images of 10-, 25-, and 50-pulse samples, magnetoresistance of the film grown on \ce{SrTiO3}.}

\bibliography{bibliography}

@article{Zhang2009,
author = {Zhang, Haijun and Liu, Chao-Xing and Qi, Xiao-Liang and Dai, Xi and Fang, Zanxi and Zhang, Shou-Cheng},
year = {2009},
month = {05},
pages = {438-442},
title = {Topological insulators in \ce{Bi2Se3}, \ce{Bi2Te3} and \ce{Sb2Te3} with a single Dirac cone on the surface},
volume = {5},
journal = {Nature Physics},
doi = {10.1038/nphys1270}}

@article{Fijalkowski2021,
author = {Fijalkowski, Kajetan and Liu, Nan and Mandal, Pankaj and Schreyeck, S. and Brunner, Karl and Gould, Charles and Molenkamp, Laurens},
year = {2021},
month = {09},
pages = {},
title = {Quantum anomalous Hall edge channels survive up to the Curie temperature},
volume = {12},
journal = {Nature Communications},
doi = {10.1038/s41467-021-25912-w}}

@article{Zhao2024,
author = {Zhao, Yi-Fan and Zhang, Ruoxi and Sun, Zi-Ting and Zhou, Ling-Jie and Zhuo, Deyi and Yan, Zi-Jie and Yi, Hemian and Wang, Ke and Chan, Moses H. W. and Liu, Chao-Xing and Law, K. T. and Chang, Cui-Zu},
title = {{3D} Quantum anomalous Hall effect in magnetic topological insulator trilayers of hundred-nanometer thickness},
journal = {Advanced Materials},
volume = {36},
number = {13},
pages = {2310249},
doi = {https://doi.org/10.1002/adma.202310249},
url = {https://advanced.onlinelibrary.wiley.com/doi/abs/10.1002/adma.202310249},
eprint = {https://advanced.onlinelibrary.wiley.com/doi/pdf/10.1002/adma.202310249},
year = {2024}}

@article{Aabdin2012,
   author = {Z. Aabdin and N. Peranio and M. Winkler and D. Bessas and J. König and R. P. Hermann and H. Böttner and O. Eibl},
   doi = {10.1007/S11664-011-1870-Z/METRICS},
   issn = {03615235},
   issue = {6},
   journal = {Journal of Electronic Materials},
   month = {6},
   pages = {1493-1497},
   publisher = {Springer},
   title = {\ce{Sb2Te3} and \ce{Bi2Te3} thin films grown by room-temperature MBE},
   volume = {41},
   url = {https://link.springer.com/article/10.1007/s11664-011-1870-z},
   year = {2012}}

@article{Krumrain2011,
   author = {J. Krumrain and G. Mussler and S. Borisova and T. Stoica and L. Plucinski and C. M. Schneider and D. Grützmacher},
   doi = {10.1016/j.jcrysgro.2011.03.008},
   issn = {00220248},
   issue = {1},
   journal = {Journal of Crystal Growth},
   month = {6},
   pages = {115-118},
   title = {MBE growth optimization of topological insulator \ce{Bi2Te3} films},
   volume = {324},
   year = {2011}}

@article{Bailini2007,
   author = {A. Bailini and F. Donati and M. Zamboni and V. Russo and M. Passoni and C. S. Casari and A. Li Bassi and C. E. Bottani},
   doi = {10.1016/j.apsusc.2007.09.039},
   issn = {01694332},
   issue = {4},
   journal = {Applied Surface Science},
   month = {12},
   pages = {1249-1254},
   publisher = {Elsevier},
   title = {Pulsed laser deposition of \ce{Bi2Te3} thermoelectric films},
   volume = {254},
   year = {2007}}

@article{Liao2019,
   author = {Zhaoliang Liao and Matthew Brahlek and Jong Mok Ok and Lauren Nuckols and Yogesh Sharma and Qiyang Lu and Yanwen Zhang and Ho Nyung Lee},
   doi = {10.1063/1.5088190},
   issn = {2166532X},
   issue = {4},
   journal = {APL Materials},
   month = {4},
   publisher = {American Institute of Physics Inc.},
   title = {Pulsed-laser epitaxy of topological insulator \ce{Bi2Te3} thin films},
   volume = {7},
   year = {2019}}

@article{Pandey2023,
   author = {Lalit Pandey and Sajid Husain and Vineet Barwal and Soumyarup Hait and Nanhe Kumar Gupta and Vireshwar Mishra and Nakul Kumar and Nikita Sharma and Dinesh Dixit and Veer Singh and Sujeet Chaudhary},
   doi = {10.1088/1361-648X/acd50a},
   issn = {1361648X},
   issue = {35},
   journal = {Journal of Physics Condensed Matter},
   month = {9},
   pmid = {37172602},
   publisher = {Institute of Physics},
   title = {Topological transport properties of highly oriented \ce{Bi2Te3} thin film deposited by sputtering},
   volume = {35},
   year = {2023}}

@article{SHANG2017532,
title = {High performance co-sputtered \ce{Bi2Te3} thin films with preferred orientation induced by \ce{MgO} substrates},
journal = {Journal of Alloys and Compounds},
volume = {726},
pages = {532-537},
year = {2017},
issn = {0925-8388},
doi = {https://doi.org/10.1016/j.jallcom.2017.07.337},
url = {https://www.sciencedirect.com/science/article/pii/S0925838817327251},
author = {Hongjing Shang and Fazhu Ding and Guicun Li and Li Wang and Fei Qu and Huiliang Zhang and Zebin Dong and He Zhang and Zhaoshun Gao and Weiwei Zhou and Hongwei Gu}}

@article{Concepcin2018,
   author = {Omar Concepción and Miguel Galván-Arellano and Vicente Torres-Costa and Aurelio Climent-Font and Daniel Bahena and Miguel Manso Silván and Arturo Escobosa and Osvaldo De Melo},
   doi = {10.1021/acs.inorgchem.8b01235},
   issn = {1520510X},
   issue = {16},
   journal = {Inorganic Chemistry},
   month = {8},
   pages = {10090-10099},
   pmid = {30066565},
   publisher = {American Chemical Society},
   title = {Controlling the epitaxial growth of \ce{Bi2Te3}, \ce{BiTe}, and \ce{Bi4Te3} pure phases by physical vapor transport},
   volume = {57},
   year = {2018}}

@article{Li2024,
author = {Li, Qile and Di Bernardo, Iolanda and Maniatis, Johnathon and McEwen, Daniel and Dominguez-Celorrio, Amelia and Bhuiyan, Mohammad T. H. and Zhao, Mengting and Tadich, Anton and Watson, Liam and Lowe, Benjamin and Vu, Thi-Hai-Yen and Trang, Chi Xuan and Hwang, Jinwoong and Mo, Sung-Kwan and Fuhrer, Michael S. and Edmonds, Mark T.},
title = {Imaging the breakdown and restoration of topological protection in magnetic topological insulator \ce{MnBi2Te4}},
journal = {Advanced Materials},
volume = {36},
number = {24},
pages = {2312004},
doi = {https://doi.org/10.1002/adma.202312004},
url = {https://advanced.onlinelibrary.wiley.com/doi/abs/10.1002/adma.202312004},
eprint = {https://advanced.onlinelibrary.wiley.com/doi/pdf/10.1002/adma.202312004},
year = {2024}}

@article{Mogi2017,
author = {Mogi, M. and Kawamura, Mayuka and Yoshimi, R. and Tsukazaki, A. and Kozuka, Y. and Shirakawa, N. and Takahashi, K. and Kawasaki, M. and Tokura, Y.},
year = {2017},
month = {02},
pages = {},
title = {A magnetic heterostructure of topological insulators as a candidate for an axion insulator},
volume = {16},
journal = {Nature Materials},
doi = {10.1038/nmat4855}}

@article{Tang2022,
   author = {Xinfeng Tang and Ziwei Li and Wei Liu and Qingjie Zhang and Ctirad Uher},
   doi = {10.1002/idm2.12009},
   issn = {2767-4401},
   issue = {1},
   journal = {Interdisciplinary Materials},
   month = {1},
   pages = {88-115},
   publisher = {Wiley},
   title = {A comprehensive review on \ce{Bi2Te3}‐based thin films: thermoelectrics and beyond},
   volume = {1},
   year = {2022}}

@article{Kriegner2017,
   author = {Dominik Kriegner and Petr Harcuba and Jozef Veselý and Andreas Lesnik and Guenther Bauer and Gunther Springholz and Václav Holý},
   doi = {10.1107/S1600576717000565},
   issn = {16005767},
   journal = {Journal of Applied Crystallography},
   pages = {369-377},
   publisher = {International Union of Crystallography},
   title = {Twin domain imaging in topological insulator \ce{Bi2Te3} and \ce{Bi2Se3} epitaxial thin films by scanning X-ray nanobeam microscopy and electron backscatter diffraction},
   volume = {50},
   year = {2017}}

@article{Mortelmans2021,
title = {Epitaxy of 2D chalcogenides: aspects and consequences of weak van der Waals coupling},
journal = {Applied Materials Today},
volume = {22},
pages = {100975},
year = {2021},
issn = {2352-9407},
doi = {https://doi.org/10.1016/j.apmt.2021.100975},
url = {https://www.sciencedirect.com/science/article/pii/S2352940721000408},
author = {Wouter Mortelmans and Stefan {De Gendt} and Marc Heyns and Clement Merckling},}

@misc{Walsh2017,
   author = {Lee A. Walsh and Christopher L. Hinkle},
   doi = {10.1016/j.apmt.2017.09.010},
   issn = {23529407},
   journal = {Applied Materials Today},
   month = {12},
   pages = {504-515},
   publisher = {Elsevier Ltd},
   title = {Van der Waals epitaxy: 2D materials and topological insulators},
   volume = {9},
   year = {2017}}

@article{Netsou2020,
author = {Netsou, Asteriona-Maria and Muzychenko, Dmitry and Dausy, Heleen and Chen, Taishi and Song, Fengqi and Schouteden, Koen and Bael, Margriet and Van Haesendonck, Chris},
year = {2020},
month = {10},
pages = {13172-13179},
title = {Identifying native point defects in the topological insulator \ce{Bi2Te3}},
volume = {14},
journal = {ACS Nano},
doi = {10.1021/acsnano.0c04861}}

@article{Pereira2021,
   author = {Vanda M. Pereira and Chi Nan Wu and Katharina Höfer and Arnold Choa and Cariad A. Knight and Jesse Swanson and Christoph Becker and Alexander C. Komarek and A. Diana Rata and Sahana Rößler and Steffen Wirth and Mengxin Guo and Minghwei Hong and Jueinai Kwo and Liu Hao Tjeng and Simone G. Altendorf},
   doi = {10.1002/pssb.202000346},
   issn = {15213951},
   issue = {1},
   journal = {Physica Status Solidi (B) Basic Research},
   month = {1},
   publisher = {Wiley-VCH Verlag},
   title = {Challenges of topological insulator research: \ce{Bi2Te3} thin films and magnetic heterostructures},
   volume = {258},
   year = {2021}}

@article{RANA2024415801,
title = {Origin of linear magnetoresistance in \ce{Bi2Te3} topological insulator: role of surface state and defects},
journal = {Physica B: Condensed Matter},
volume = {679},
pages = {415801},
year = {2024},
issn = {0921-4526},
doi = {https://doi.org/10.1016/j.physb.2024.415801},
url = {https://www.sciencedirect.com/science/article/pii/S092145262400142X},
author = {Nabakumar Rana and Pintu Singha and Suchandra Mukherjee and Subarna Das and Gangadhar Das and Apurba Kanti Deb and Sujay Chakravarty and S. Bandyopadhyay and Aritra Banerjee}}

@Article{Necas2012,
author = {Nečas, David and Klapetek, Petr},
affiliation = {CEITEC — Central European Institute of Technology, Masaryk University Kamenice 753/5, 625 00 Brno, Czech Republic},
title = {Gwyddion: an open-source software for {SPM} data analysis},
journal = {Central European Journal of Physics},
publisher = {Versita, co-published with Springer-Verlag GmbH},
issn = {1895-1082},
pages = {181-188},
volume = {10},
issue = {1},
year = {2012},
doi = {10.2478/s11534-011-0096-2}}

@article{Vignaud:rg5154,
author = "Vignaud, Guillaume and Gibaud, Alain",
title = "{{\it REFLEX}: a program for the analysis of specular X-ray and neutron reflectivity data}",
journal = "Journal of Applied Crystallography",
year = "2019",
volume = "52",
number = "1",
pages = "201--213",
month = "Feb",
doi = {10.1107/S1600576718018186},
url = {https://doi.org/10.1107/S1600576718018186}}

@inproceedings{Russo2008,
   author = {V. Russo and A. Bailini and M. Zamboni and M. Passoni and C. Conti and C. S. Casari and A. Li Bassi and C. E. Bottani},
   doi = {10.1002/jrs.1874},
   issn = {10974555},
   issue = {2},
   booktitle = {Journal of Raman Spectroscopy},
   pages = {205-210},
   publisher = {John Wiley and Sons Ltd},
   title = {Raman spectroscopy of Bi-Te thin films},
   volume = {39},
   year = {2008}}

@article{KUZNETSOV2016122,
title = {Growth of \ce{Bi2Te3} films and other phases of Bi-Te system by MOVPE},
journal = {Journal of Crystal Growth},
volume = {455},
pages = {122-128},
year = {2016},
issn = {0022-0248},
doi = {https://doi.org/10.1016/j.jcrysgro.2016.09.055},
url = {https://www.sciencedirect.com/science/article/pii/S0022024816305589},
author = {P.I. Kuznetsov and V.O. Yapaskurt and B.S. Shchamkhalova and V.D. Shcherbakov and G.G. Yakushcheva and V.A. Luzanov and V.A. Jitov},}

@article{Xu2015Raman,
    author = {Xu, Hao and Song, Yuxin and Pan, Wenwu and Chen, Qimiao and Wu, Xiaoyan and Lu, Pengfei and Gong, Qian and Wang, Shumin},
    title = {Vibrational properties of epitaxial \ce{Bi4Te3} films as studied by Raman spectroscopy},
    journal = {AIP Advances},
    volume = {5},
    number = {8},
    pages = {087103},
    year = {2015},
    month = {08},
    issn = {2158-3226},
    doi = {10.1063/1.4928217},
    url = {https://doi.org/10.1063/1.4928217},
    eprint = {https://pubs.aip.org/aip/adv/article-pdf/doi/10.1063/1.4928217/8442466/087103_1_online.pdf}}

@article{rodrigues2023,
    author = {Rodrigues, Leonarde N. and de Araujo, C. I. L. and Mello, S. L. A. and Laverock, J. and Fonseca, Jakson M. and Schwarzacher, W. and Inoch, Wesley F. and Ferreira, Sukarno O.},
    title = {Growth of \ce{Bi2Te3} topological insulator ultra-thin layers via molecular beam epitaxy on \ce{GaAs} (100)},
    journal = {Journal of Applied Physics},
    volume = {134},
    number = {8},
    pages = {085301},
    year = {2023},
    month = {08},
    issn = {0021-8979},
    doi = {10.1063/5.0155332},
    url = {https://doi.org/10.1063/5.0155332},
    eprint = {https://pubs.aip.org/aip/jap/article-pdf/doi/10.1063/5.0155332/18093215/085301\_1\_5.0155332.pdf},}

@article{Shahil2010,
   author = {K. M.F. Shahil and M. Z. Hossain and D. Teweldebrhan and A. A. Balandin},
   doi = {10.1063/1.3396190/338640},
   issn = {00036951},
   issue = {15},
   journal = {Applied Physics Letters},
   month = {4},
   pages = {153103},
   publisher = {AIP Publishing},
   title = {Crystal symmetry breaking in few-quintuple \ce{Bi2Te3} films: applications in nanometrology of topological insulators},
   volume = {96},
   url = {/aip/apl/article/96/15/153103/338640/Crystal-symmetry-breaking-in-few-quintuple-Bi2Te3},
   year = {2010}}

@article{Liu2012,
   author = {Y. Liu and M. Weinert and L. Li},
   doi = {10.1103/PhysRevLett.108.115501},
   issn = {00319007},
   issue = {11},
   journal = {Physical Review Letters},
   month = {3},
   title = {Spiral growth without dislocations: Molecular beam epitaxy of the topological insulator \ce{Bi2Se3} on epitaxial graphene/\ce{SiC}(0001)},
   volume = {108},
   year = {2012}}

@article{Ferhat2000,
title = {Mechanisms of spiral growth in \ce{Bi2Te3} thin films grown by the hot-wall-epitaxy technique},
journal = {Journal of Crystal Growth},
volume = {218},
number = {2},
pages = {250-258},
year = {2000},
issn = {0022-0248},
doi = {https://doi.org/10.1016/S0022-0248(00)00582-0},
url = {https://www.sciencedirect.com/science/article/pii/S0022024800005820},
author = {Marhoun Ferhat and Jean Claude Tedenac and Jiro Nagao}}

@Article{Mortelmans2020,
author={Mortelmans, Wouter and Nalin Mehta, Ankit and Balaji, Yashwanth and Sergeant, Stefanie and Meng, Ruishen and Houssa, Michel and De Gendt, Stefan and Heyns, Marc and Merckling, Clement},
title={On the van der Waals epitaxy of homo-/heterostructures of transition metal dichalcogenides},
journal={ACS Applied Materials {\&} Interfaces},
year={2020},
month={Jun},
day={17},
publisher={American Chemical Society},
volume={12},
number={24},
pages={27508-27517},
issn={1944-8244},
doi={10.1021/acsami.0c05872},
url={https://doi.org/10.1021/acsami.0c05872}}

@article{analytis2010bulk,
  title={Bulk Fermi surface coexistence with Dirac surface state in \ce{Bi2Se3}: a comparison of photoemission and Shubnikov-de Haas measurements},
  author={Analytis, James G and Chu, Jiun-Haw and Chen, Yulin and Corredor, Felipe and McDonald, Ross D and Shen, ZX and Fisher, Ian R},
  journal={Physical Review B—Condensed Matter and Materials Physics},
  volume={81},
  number={20},
  pages={205407},
  year={2010},
  publisher={APS}}

@article{ngabonziza2022quantum,
  title={Quantum transport and potential of topological states for thermoelectricity in \ce{Bi2Te3} thin films},
  author={Ngabonziza, Prosper},
  journal={Nanotechnology},
  volume={33},
  number={19},
  pages={192001},
  year={2022},
  publisher={IOP Publishing}}

@inproceedings{lu2014weak,
  title={Weak localization and weak anti-localization in topological insulators},
  author={Lu, Hai-Zhou and Shen, Shun-Qing},
  booktitle={Spintronics Vii},
  volume={9167},
  pages={263--273},
  year={2014},
  organization={SPIE}}

@article{he2011impurity,
  title={Impurity effect on weak antilocalization in the topological insulator \ce{Bi2Te3}},
  author={He, Hong-Tao and Wang, Gan and Zhang, Tao and Sou, Iam-Keong and Wong, George K L and Wang, Jian-Nong and Lu, Hai-Zhou and Shen, Shun-Qing and Zhang, Fu-Chun},
  journal={Physical review letters},
  volume={106},
  number={16},
  pages={166805},
  year={2011},
  publisher={APS}}

@article{abramowitz1965formulas,
  title={Handbook of mathematical functions with formulas, graphs, and mathematical tables},
  author={Abramowitz, Milton and Stegun, Irene A},
  journal={National Bureau of Standards Applied Mathematics Series. e},
  volume={55},
  pages={953},
  year={1965}}

@article{hikami1980spin,
  title={Spin-orbit interaction and magnetoresistance in the two dimensional random system},
  author={Hikami, Shinobu and Larkin, Anatoly I and Nagaoka, Yosuke},
  journal={Progress of Theoretical Physics},
  volume={63},
  number={2},
  pages={707--710},
  year={1980},
  publisher={Oxford University Press}}

@article{chiu2013weak,
  title={Weak antilocalization in topological insulator \ce{Bi2Te3} microflakes},
  author={Chiu, Shao-Pin and Lin, Juhn-Jong},
  journal={Physical Review B—Condensed Matter and Materials Physics},
  volume={87},
  number={3},
  pages={035122},
  year={2013},
  publisher={APS}}

@article{steinberg2011electrically,
  title={Electrically tunable surface-to-bulk coherent coupling in topological insulator thin films},
  author={Steinberg, Hadar and Lalo{\"e}, J-B and Fatemi, Valla and Moodera, Jagadeesh S and Jarillo-Herrero, Pablo},
  journal={Physical Review B—Condensed Matter and Materials Physics},
  volume={84},
  number={23},
  pages={233101},
  year={2011},
  publisher={APS}}

@article{chen2010gate,
  title={Gate-voltage control of chemical potential and weak antilocalization in \ce{Bi2Se3}},
  author={Chen, J and Qin, HJ and Yang, F and Liu, J and Guan, T and Qu, FM and Zhang, GH and Shi, JR and Xie, XC and Yang, CL and others},
  journal={Physical Review Letters},
  volume={105},
  number={17},
  pages={176602},
  year={2010},
  publisher={APS}}

@article{PhysRevLett.109.066803,
  title = {Manifestation of topological protection in transport properties of epitaxial \ce{Bi2Se3} thin films},
  author = {Taskin, A. A. and Sasaki, Satoshi and Segawa, Kouji and Ando, Yoichi},
  journal = {Phys. Rev. Lett.},
  volume = {109},
  issue = {6},
  pages = {066803},
  numpages = {5},
  year = {2012},
  month = {Aug},
  publisher = {American Physical Society},
  doi = {10.1103/PhysRevLett.109.066803},
  url = {https://link.aps.org/doi/10.1103/PhysRevLett.109.066803}}

@article{barthel2008dephasing,
  title={Dephasing and the steady state in quantum many-particle systems},
  author={Barthel, Thomas and Schollw{\"o}ck, Ulrich},
  journal={Physical review letters},
  volume={100},
  number={10},
  pages={100601},
  year={2008},
  publisher={APS}}

@article{lin2002recent,
  title={Recent experimental studies of electron dephasing in metal and semiconductor mesoscopic structures},
  author={Lin, Juhn-Jong and Bird, JP},
  journal={Journal of Physics: Condensed Matter},
  volume={14},
  number={18},
  pages={R501--R596},
  year={2002}}

@book{efros2012electron,
  title={Electron-electron interactions in disordered systems},
  author={Efros, Alex L and Pollak, Michael},
  volume={10},
  year={2012},
  publisher={Elsevier}}

\end{document}


\maketitle

\begin{figure}[h]
  \centering
  \includegraphics[width=0.75\linewidth]{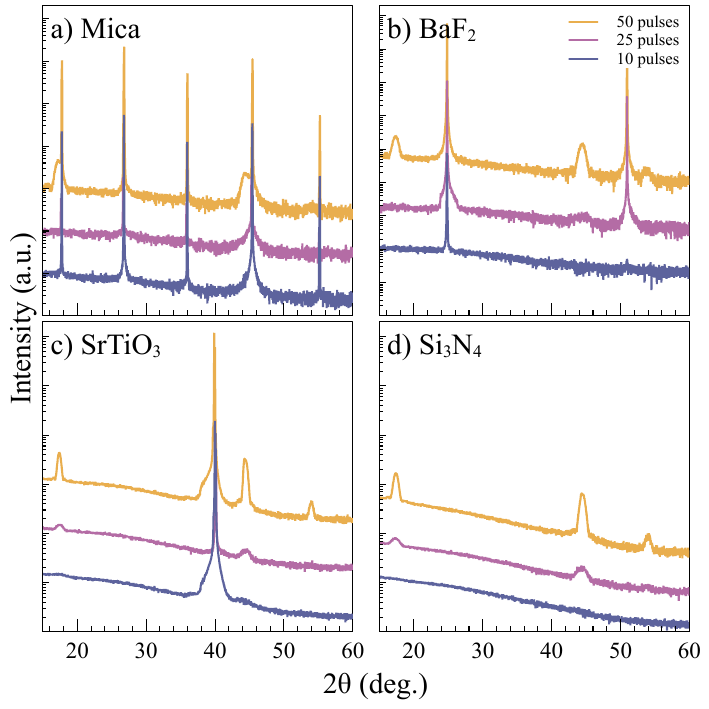}
  \caption{Widescan Bragg-Brentano diffractograms of deposited \ce{Bi2Te3} thin films.}
  \label{si1}
\end{figure}

\begin{figure}[h]
  \centering
  \includegraphics[width=0.75\linewidth]{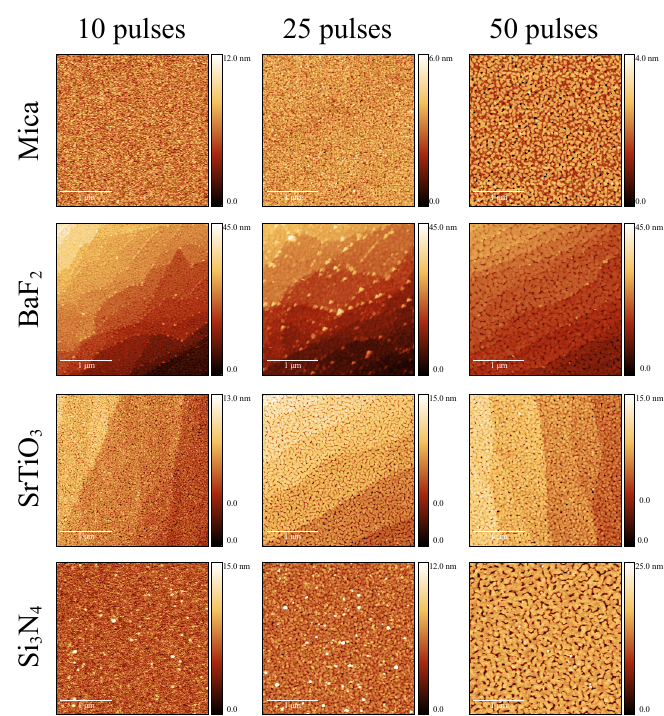}
  \caption{AFM scans of the 10-, 25- and 50-pulse \ce{Bi2Te3} thin films.}
  \label{si2}
\end{figure}

\begin{figure}[h]
  \centering
  \includegraphics[width=0.75\linewidth]{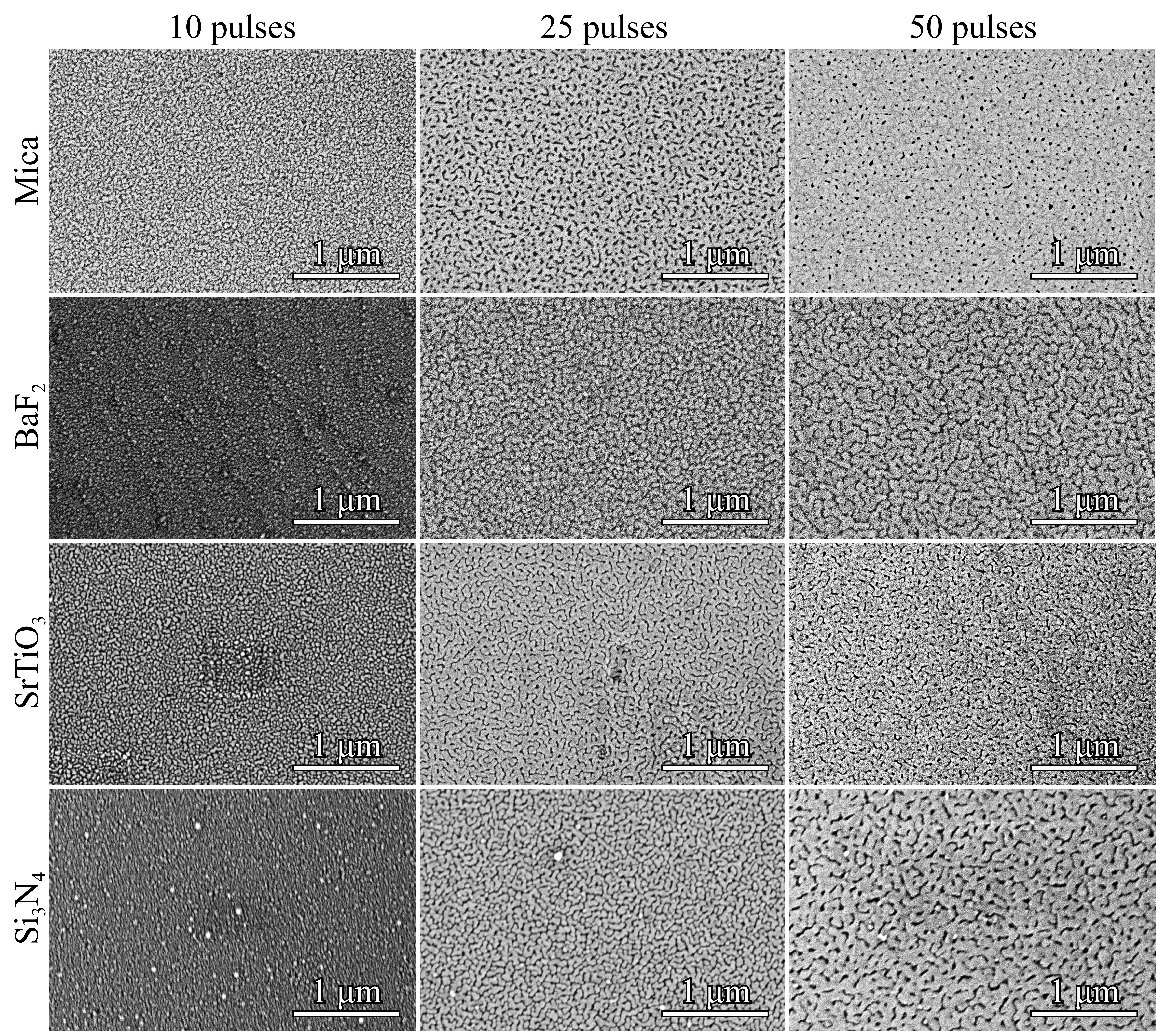}
  \caption{SEM images of the 10-, 25-, and 50-pulse \ce{Bi2Te3} thin films.}
  \label{si3}
\end{figure}

\begin{figure}[h]
  \centering
  \includegraphics[width=0.75\linewidth]{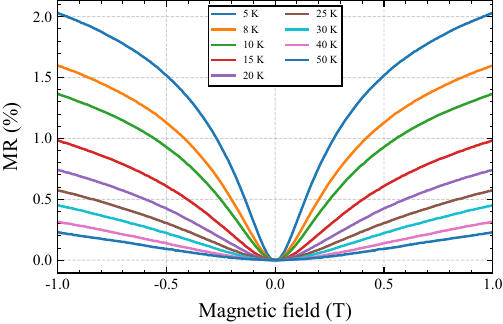}
  \caption{Magnetoresistivity of the \ce{Bi2Te3} thin film deposited on the \ce{SrTiO3} substrate.}
  \label{si4}
\end{figure}

\begin{table}
    \caption{Calculated ratio of $E^2_g$ to $A^2_{1g}$ Raman peaks of the 10-, 25-, and 50-pulse \ce{Bi2Te3} thin films}
    \centering
    \begin{tabular}{l|c|c|c|c}
    \hline
    No. of pulses & Mica & BaF$_{2}$ & SrTiO$_{3}$  & Si$_{3}$N$_{4}$ \\
    \hline \hline
    10 & 2.65 & 1.95 & --- & 2.25 \\
    25 & 2.78 & 2.80 & 3.53 & 2.92 \\
    50 & 3.05 & 3.31 & 3.36 & 3.35 \\
    \hline
    \end{tabular}
    \label{tabsi1}
\end{table}

\begin{table}
  \caption{The summary of topographical features of \ce{Bi2Te3} films across the selected numbers for laser pulses}
  \centering
  \label{tabsi2}
    \begin{tabular}{l|l|l|l|l|l|l|l|l|l|l|l|l}
        \hline
        & \multicolumn{3}{| c |}{\text{(001)Mica}} & \multicolumn{3}{| c |}{\text{(111)BaF$_{2}$}} & \multicolumn{3}{| c |}{\text{(111)SrTiO$_{3}$}}  & \multicolumn{3}{| c }{\text{Si$_{3}$N$_{4}$}} \\
        \hline 
             No. of pulses & 10 & 25 & 50 & 10 & 25 & 50  & 10 & 25 & 50  & 10 & 25 & 50 \\
        \hline \hline
            \text{thickness (nm)} & 1 & 3 & 8 & 4 & 7 & 9  & 7 & 11 & 16  & 4 & 8 & 12\\
            \text{fraction coverage (\%)} & 80 & 86 & 98 & 76 & 81 & 89  & 87 & 90 & 95  & 60 & 72 & 91\\
            \text{formula units per area (nm$^{-2}$)}& 5 & 15 & 46 & 18 & 33 & 47  & 36 & 58 & 90 & 14 & 34 & 64 \\
            \text{roughness (nm)} & 1.3 & 0.5 & 0.7 & 1.5 & 1.8 & 2.1  & 1.1 & 1.7 & 1.3  & 1.6 & 1.1 & 3.4\\
        \hline
  \end{tabular}
\end{table}